\documentclass[floatfix,showpacs,prl,twocolumn]{revtex4-1}
\usepackage{makeidx}
\usepackage{eurosym}
\usepackage{graphicx}
\usepackage{graphics}
\usepackage{amssymb}
\usepackage{amsmath}
\usepackage{color}
\usepackage{float}

\begin{document}

\title{2D Weyl Fermi gas model of Superconductivity in the Surface state of
a Topological Insulator at High Magnetic fields}
\author{Vladimir Zhuravlev}
\author{Wenye Duan}
\author{Tsofar Maniv}
\email{e-mail:maniv@tx.technion.ac.il}
\date{\today }

\begin{abstract}
The Nambu-Gorkov Green's function approach is applied to strongly type-II
superconductivity in a 2D spin-momentum locked (Weyl) Fermi gas model at
high perpendicular magnetic fields. When the chemical potential is
sufficiently close to the branching (Dirac) point, such that the cyclotron
effective mass, $m^{\ast }$, is a very small fraction of the free electron
mass, $m_{e}$, relatively large portion of the $H-T$ phase diagram is
exposed to magneto-quantum oscillation effects. This model system is
realized in the 2D superconducting state, observed recently on the surface
of the topological insulator Sb$_{2}$Te$_{3} $, for which high field
measurements were reported at low carrier densities with $m^{\ast
}=0.065m_{e}$. Calculations of the pairing condensation energy in such a
system, as a function of $H$ and $T$, using both the Weyl model and a
reference standard model, that exploits a simple quadratic dispersion law,
are found to yield indistinguishable results in comparison with the
experimental data. \ Significant deviations from the predictions of the
standard model are found only for very small carrier densities, when the
cyclotron energy becomes very large, the Landau level filling factors are
smaller than unity, and the Fermi energy shrinks below the cutoff energy.
\end{abstract}

\maketitle

\affiliation{Schulich Faculty of Chemistry, Technion-Israel Institute of Technology,
	Haifa 32000, Israel }

\affiliation{School of Physics, Peking University, Beijing 100871, China }

\affiliation{Schulich Faculty of Chemistry, Technion-Israel Institute of Technology,
	Haifa 32000, Israel}

\section{Introduction\protect \bigskip}

{The recent discoveries of surface and interface superconductivity with} {%
exceptionally high superconducting (SC) transition temperatures in several
material structures} \cite{BosovicNphys14},\cite{FengGeNmat15}, \cite%
{MannaArXive16} have drawn much attention to  the phenomenon of strong
type-II superconductivity in two-dimensional (2D) electron systems, in which
the application of high magnetic fields can lead to exotic phenomena both in
the normal and SC states\cite{Maniv01}.\ {Of special interest is the unique
situation of the 2D superconductivity realized in surface states of
topological insulators, e.g. }Sb$_{2}$Te$_{3}$ \cite{Zhao15}, {where the
chemical potential} $E_{F}$ {is close to a Dirac point }\cite{Zhang09}{\
(with Fermi velocity }$v${) and the cyclotron effective mass}, $m^{\ast
}=E_{F}/v^{2}$\cite{Katsnelson12} {is a small fraction (e.g. 0.065 in }Sb$%
_{2}$Te$_{3}$, see also \cite{Arnold16}{) of the free electron mass} $m_{e}$%
, {resulting in a dramatic enhancement of the cyclotron frequency}, $\omega
_{c}=eH/m^{\ast }c$, {and the corresponding Landau level (LL) energy
spacing. In a recent paper \cite{ZDMPRB17} we have exploited a standard
electron gas model, with a quadratic energy-momentum dispersion and an
effective band mass }$m^{\ast }={0.065}m_{e}$, in a systematic investigation
of the quasi-particle states and the SC pair-potential in the vortex lattice
state of this system under high perpendicular magnetic fields,{\ by solving
self-consistently the corresponding }Bogoliubov de Gennes (BdG) equations.
The results account reasonably well for the 2D SC state observed on the
surface of Sb$_{2}$Te$_{3}$ under magnetic fields of up to 3 T \cite{Zhao15}%
, revealing a strong type-II superconductivity at unusually low carrier
density and small cyclotron effective mass, which can be realized only in
the strong coupling ($\lambda \sim 1$) superconductor limit. {This unique
situation is due to the proximity of the Fermi energy to a Dirac point,
which implies that other materials in the emerging field of surface
superconductivity, with metallic surface states and Dirac dispersion law
around the Fermi energy, can show similar features. }

It should be noted, however, that the use of the standard LL spectrum,
arising from a parabolic band-structure, in the self-consistent BdG theory,
presented in Ref.{\cite{ZDMPRB17}, has been done heuristically, without
actual derivation from} the effective 2D Weyl Hamiltonian describing the
helical surface states {observed in these topological insulators  \cite%
{Tkachov16}, }\cite{Zhao15}. Such a derivation is particularly necessary for
the spin-momentum locked model under study here, since SC pairing involves
certain spin-orbital correlations. 

Our purpose in the present paper is, therefore, two-fold: First, to develop
the formal framework for solving the self consistency equation for the SC
order parameter in the 2D Weyl model Hamiltonian under a strong
perpendicular magnetic field, and then exploit the developed formalism in a
study of the transition to superconductivity in comparison with the well
known results of the standard model\cite{Maniv01}. The SC transition in
helical surface states of topological insulators, such as those reported,
e.g. in Ref.\cite{Zhao15}, is then comparatively studied with respect to
both models. It is found that, similar to the well known solution of the
linearized self consistency equation, derived by Helfand and Werthamer for
the standard model \cite{HW64}, {\cite{ZDMPRB17}} the desired solution of
the corresponding integral equation in the 2D Weyl model is greatly
facilitated by initially finding analytical solutions of the eigenvalue
equation for the SC order parameter. \ Furthermore, the calculated $H-T$
phase diagram for the Weyl model in the semiclassical limit (i.e. for LL
filling factors $n_{F}>1$) can be directly mapped onto that found for the
standard model, having the same Fermi surface parameters $E_{F}$ and $v$,
and a cyclotron effective mass equal to $m^{\ast }=E_{F}/2v^{2}$.
Significant deviations from the predicted mapping are found only for very
small carrier densities, when the cyclotron energy becomes very large, the
LL filling factors are smaller than unity, and the Fermi energy shrinks
below the cutoff energy.

\section{The 2D spin-momentum locked (Weyl) Fermion gas model}

To describe the underlying normal surface state electron, with charge $-e$ ,
in a topological insulator under a magnetic field $\mathbf{H=}\left(
0,0,H\right) $ (vector potential in the Landau gauge, $\mathbf{A=}\left(
-Hy,0,0\right) $), we exploit the Weyl Hamiltonian: \cite{Tkachov16}

\begin{equation}
\widehat{h}\left( \mathbf{r}\right) =\hbar v\left( \widehat{\sigma }_{x}%
\widetilde{p}_{x}+\widehat{\sigma }_{y}\widetilde{p}_{y}\right) -E_{F}%
\widehat{\sigma }_{0}  \label{WeylH}
\end{equation}

with the Pauli matrices: $\  \widehat{\sigma }_{x}=\left( 
\begin{array}{cc}
0 & 1 \\ 
1 & 0%
\end{array}%
\right) ;\widehat{\sigma }_{y}=\left( 
\begin{array}{cc}
0 & -i \\ 
i & 0%
\end{array}%
\right) ;\widehat{\sigma }_{0}=\left( 
\begin{array}{cc}
1 & 0 \\ 
0 & 1%
\end{array}%
\right) $, and the gauge invariant momentum $\widetilde{\mathbf{p}}\mathbf{%
\equiv }\left( -i \mathbf{\nabla +}\left(e/\hbar c\right)\mathbf{A}\right) $%
, such that: 
\begin{widetext}
\begin{equation}
\widehat{h}\left( \mathbf{r}\right) =\left( 
\begin{array}{cc}
0 & -\hbar v\frac{\partial }{\partial y}-\hbar v\left( i\frac{\partial }{%
\partial x}+\frac{y}{a_{H}^{2}}\right) \\ 
\hbar v\frac{\partial }{\partial y}-\hbar v\left( i\frac{\partial }{\partial
x}+\frac{y}{a_{H}^{2}}\right) & 0%
\end{array}%
\right) .
\end{equation}
In these equations $v$ is the Fermi velocity and $a_{H}\equiv \sqrt{\frac{%
\hbar c}{eH}}$ is the magnetic length. Note that Zeeman spin-splitting is
neglected with respect to the cyclotron energy in Eq.\ref{WeylH} due to the
very small cyclotron effective mass considered here.

The corresponding Weyl equation for the spinor $\left( 
\begin{array}{c}
\psi _{\uparrow }\left( \mathbf{r}\right) \\ 
\psi _{\downarrow }\left( \mathbf{r}\right)%
\end{array}%
\right) $ takes the form: 
\begin{equation}
\  \left( 
\begin{array}{cc}
-E_{F} & -\hbar v\frac{\partial }{\partial y}-\hbar v\left( i\frac{\partial 
}{\partial x}+\frac{y}{a_{H}^{2}}\right) \\ 
\hbar v\frac{\partial }{\partial y}-\hbar v\left( i\frac{\partial }{\partial
x}+\frac{y}{a_{H}^{2}}\right) & -E_{F}%
\end{array}%
\right) \left( 
\begin{array}{c}
\psi _{\uparrow }\left( \mathbf{r}\right) \\ 
\psi _{\downarrow }\left( \mathbf{r}\right)%
\end{array}%
\right) =E\left( 
\begin{array}{c}
\psi _{\uparrow }\left( \mathbf{r}\right) \\ 
\psi _{\downarrow }\left( \mathbf{r}\right)%
\end{array}%
\right)  \label{WeylEq}
\end{equation}%
\bigskip

Expressing all length variables in units of the magnetic length $a_{H}$, and
introducing the dimensionless energy variable $\mu \equiv \frac{E_{F}a_{H}}{%
\sqrt{2}\hbar v}$, where all other energy symbols refer in what follows to
quantities measured in units of the cyclotron energy $\hbar \omega
_{c}\equiv \frac{\sqrt{2}\hbar v}{a_{H}}$, we write the corresponding
mean-field Hamiltonian for singlet pairing in Nambu representation:

\begin{equation}
\widehat{H}=\left( 
\begin{array}{cc}
\frac{1}{\sqrt{2}}\widehat{\boldsymbol{\mathbf{\sigma }}}\cdot \widetilde{%
\mathbf{p}}-\mu & i\widehat{\sigma }_{y}\Delta \left( \mathbf{r}\right) \\ 
-i\widehat{\sigma }_{y}\Delta ^{\ast }\left( \mathbf{r}\right) & -\frac{1}{%
\sqrt{2}} \widehat{\boldsymbol{\mathbf{\sigma }}}^{\ast }\cdot \widetilde{%
\mathbf{p}}^{\ast }+\mu%
\end{array}%
\right)  \label{NambuH}
\end{equation}%
where the spin-singlet order parameter is defined by: $\Delta ^{\ast }\left( 
\mathbf{r}\right) \equiv -\left(\left \vert V\right \vert /\hbar \omega_{c}
\right) \left \langle \psi _{\downarrow }^{\dagger }\left( \mathbf{r}\right)
\psi _{\uparrow }^{\dagger }\left( \mathbf{r}\right) \right \rangle \equiv
\Delta _{\uparrow \downarrow }^{\ast }\left( \mathbf{r}\right) =-\Delta
_{\downarrow \uparrow }^{\ast }\left( \mathbf{r}\right) $ \cite%
{InducedTriplet} and $\widehat{\boldsymbol{\mathbf{\sigma }}}\equiv
\left(\sigma_{x},\sigma_{y}\right)$.

The corresponding Nambu field operators:%
\begin{equation*}
\widehat{\Psi }\left( \mathbf{r};t\right) \equiv \left[ 
\begin{array}{c}
\psi _{\uparrow }\left( \mathbf{r};t\right) \\ 
\psi _{\downarrow }\left( \mathbf{r};t\right) \\ 
\psi _{\uparrow }^{\dagger }\left( \mathbf{r};t\right) \\ 
\psi _{\downarrow }^{\dagger }\left( \mathbf{r};t\right)%
\end{array}%
\right] ,\widehat{\Psi }^{\dagger }\left( \mathbf{r};t\right) \equiv \left[ 
\begin{array}{cccc}
\psi _{\uparrow }^{\dagger }\left( \mathbf{r};t\right) & \psi _{\downarrow
}^{\dagger }\left( \mathbf{r};t\right) & \psi _{\uparrow }\left( \mathbf{r}%
;t\right) & \psi _{\downarrow }\left( \mathbf{r};t\right)%
\end{array}%
\right]
\end{equation*}%
satisfy the equation of motion $i\partial _{t}\widehat{\Psi }\left( \mathbf{r%
};t\right) =\widehat{H}\widehat{\Psi }\left( \mathbf{r};t\right) $ ,
resulting in the corresponding equations for the Nambu-Gorkov time-ordered
Green's functions $4\times 4$ matrix, $\widehat{G}\left( \mathbf{r,r}%
^{\prime };t-t^{\prime }\right) \equiv -i\left \langle T\widehat{\Psi }%
\left( \mathbf{r};t\right) \widehat{\Psi }^{\dagger }\left( \mathbf{r}%
^{\prime };t^{\prime }\right) \right \rangle $:

\bigskip

\begin{equation}
\left[ i\partial _{t}-\left( 
\begin{array}{cc}
\frac{1}{\sqrt{2}}\widehat{\mathbf{\sigma }}\cdot \widetilde{\mathbf{p}}-\mu
& i\widehat{\sigma }_{y}\Delta ^{\ast }\left( \mathbf{r}\right) \\ 
-i\widehat{\sigma }_{y}\Delta ^{\ast }\left( \mathbf{r}\right) & -\frac{1}{%
\sqrt{2}}\widehat{\mathbf{\sigma }}^{\ast }\cdot \widetilde{\mathbf{p}}%
^{\ast }+\mu%
\end{array}%
\right) \right] \left( 
\begin{array}{cc}
\widehat{G}_{11}\left( \mathbf{r,r}^{\prime };t-t^{\prime }\right) & 
\widehat{G}_{12}\left( \mathbf{r,r}^{\prime };t-t^{\prime }\right) \\ 
\widehat{G}_{21}\left( \mathbf{r,r}^{\prime };t-t^{\prime }\right) & 
\widehat{G}_{22}\left( \mathbf{r,r}^{\prime };t-t^{\prime }\right)%
\end{array}%
\right) =\delta \left( t-t^{\prime }\right) \delta \left( \mathbf{r}-\mathbf{%
r}^{\prime }\right)  \label{DiffG}
\end{equation}

\bigskip

Time-Fourier transforming with frequency $\omega $ and rewriting Eq.\ref%
{DiffG} in its integral form, the relevant parts of these equations for our
purpose here is written in the form:

\bigskip

\begin{eqnarray}
\widehat{G}_{11}\left( \mathbf{r,r}^{\prime };\omega \right) &=&\widehat{G}%
_{11}^{\left( 0\right) }\left( \mathbf{r,r}^{\prime };\omega \right) +\int d%
\mathbf{r}^{\prime \prime }\widehat{G}_{11}^{\left( 0\right) }\left( \mathbf{%
r,r}^{\prime \prime };\omega \right) i\widehat{\sigma }_{y}\Delta \left( 
\mathbf{r}^{\prime \prime }\right) \widehat{G}_{21}\left( \mathbf{r}^{\prime
\prime }\mathbf{,r}^{\prime };\omega \right) ,  \label{IntG11G21} \\
\widehat{G}_{21}\left( \mathbf{r,r}^{\prime };\omega \right) &=&\int d%
\mathbf{r}^{\prime \prime }\widehat{G}_{11}^{\left( 0\right) T}\left( 
\mathbf{r}^{\prime \prime }\mathbf{,r};-\omega \right) i\widehat{\sigma }%
_{y}\Delta ^{\ast }\left( \mathbf{r}^{\prime \prime }\right) \widehat{G}%
_{11}\left( \mathbf{r}^{\prime \prime }\mathbf{,r}^{\prime };\omega \right)
\label{IntG21G11}
\end{eqnarray}%
where the upper-left block of the Normal state $2\times 2$ Green's function
matrix: $\widehat{G}_{11}^{\left( 0\right) }\left( \mathbf{r,r}^{\prime
};\omega \right) \equiv \left( 
\begin{array}{cc}
G_{\uparrow \uparrow }^{\left( 0\right) }\left( \mathbf{r,r}^{\prime
};\omega \right) & G_{\uparrow \downarrow }^{\left( 0\right) }\left( \mathbf{%
r,r}^{\prime };\omega \right) \\ 
G_{\downarrow \uparrow }^{\left( 0\right) }\left( \mathbf{r,r}^{\prime
};\omega \right) & G_{\downarrow \downarrow }^{\left( 0\right) }\left( 
\mathbf{r,r}^{\prime };\omega \right)%
\end{array}%
\right) $, satisfies the equation:

\begin{equation}
\left( \omega +\mu -\frac{1}{\sqrt{2}}\widehat{\mathbf{\sigma }}\cdot 
\widetilde{\mathbf{p}}\right) \widehat{G}_{11}^{\left( 0\right) }\left( 
\mathbf{r,r}^{\prime };\omega \right) =\delta \left( \mathbf{r}-\mathbf{r}%
^{\prime }\right)  \label{DiffGN11}
\end{equation}%
and its transpose with frequency $-\omega $, $\widehat{G}_{11}^{\left(
0\right) T}\left( \mathbf{r}^{\prime }\mathbf{,r};-\omega \right) $,
satisfies the dual equation:

\begin{equation}
\left( \omega -\mu +\frac{1}{\sqrt{2}}\mathbf{\sigma }^{\ast }\cdot 
\widetilde{\mathbf{p}}^{\ast }\right) \widehat{G}_{11}^{\left( 0\right)
T}\left( \mathbf{r}^{\prime }\mathbf{,r};-\omega \right) =-\delta \left( 
\mathbf{r}-\mathbf{r}^{\prime }\right)  \label{DiffGN11T}
\end{equation}

\bigskip

Expanding the above normal state Green's functions in terms of the complete
set of solutions, $\varphi _{n}\left( y-k_{x}\right) =\frac{1}{\pi ^{1/4}%
\sqrt{2^{n}n!}}e^{-\frac{1}{2}\left( y-k_{x}\right) ^{2}}H_{n}\left(
y-k_{x}\right) $ , of the eigenstate equation:$\frac{1}{2}\left[ -\partial
_{y}^{2}+\left( y-k_{x}\right) ^{2}-1\right] \varphi _{n}\left(
y-k_{x}\right) =n\varphi _{n}\left( y-k_{x}\right) $, where $H_{n}\left(
y\right) $ is Hermite polynomial of order $n=0,1,2,...,$ we find:

\begin{equation}
G_{\uparrow \uparrow }^{\left( 0\right) }\left( \mathbf{r,r}^{\prime
};\omega \right) = \frac{1}{L_{x}}\sum \limits_{k_{x}}e^{ik_{x}\left(
x-x^{\prime }\right) }\sum \limits_{n=0}^{\infty }\frac{\left( \omega +\mu
\right) \varphi _{n}\left( y-k_{x}\right) \varphi _{n}\left( y^{\prime
}-k_{x}\right) }{\left( \omega +\mu \right) ^{2}-\left( n+1\right) }
\label{gUU}
\end{equation}

\bigskip

\begin{equation}
G_{\uparrow \downarrow }^{\left( 0\right) }\left( \mathbf{r,r}^{\prime
};\omega \right) = \frac{1}{L_{x}}\sum \limits_{k_{x}}e^{ik_{x}\left(
x-x^{\prime }\right) }\sum \limits_{n=0}^{\infty }\frac{\left( -\sqrt{n}%
\right) \varphi _{n-1}\left( y-k_{x}\right) \varphi _{n}\left( y^{\prime
}-k_{x}\right) }{\left( \omega +\mu \right) ^{2}-n}  \label{gUD}
\end{equation}

\bigskip

\begin{equation}
G_{\downarrow \uparrow }^{\left( 0\right) }\left( \mathbf{r,r}^{\prime
};\omega \right) = \frac{1}{L_{x}}\sum \limits_{k_{x}}e^{ik_{x}\left(
x-x^{\prime }\right) }\sum \limits_{n=0}^{\infty }\frac{\left( -\sqrt{n+1}%
\right) \varphi _{n+1}\left( y-k_{x}\right) \varphi _{n}\left( y^{\prime
}-k_{x}\right) }{\left( \omega +\mu \right) ^{2}-\left( n+1\right) }
\label{gDU}
\end{equation}

\bigskip

\begin{equation}
G_{\downarrow \downarrow }^{\left( 0\right) }\left( \mathbf{r,r}^{\prime
};\omega \right) = \frac{1}{L_{x}}\sum \limits_{k_{x}}e^{ik_{x}\left(
x-x^{\prime }\right) }\sum \limits_{n=0}^{\infty }\frac{\left( \omega +\mu
\right) \varphi _{n}\left( y-k_{x}\right) \varphi _{n}\left( y^{\prime
}-k_{x}\right) }{\left( \omega +\mu \right) ^{2}-n}  \label{gDD}
\end{equation}%
where $L_{x}$ is the surface size along the x-axis (measured in units of $%
a_{H}$).

Leading order expansion of the integral equations Eq.\ref{IntG21G11} in the
order parameter $\Delta \left( \mathbf{r}\right) $ yields:

\bigskip

\begin{equation*}
\widehat{G}_{21}\left( \mathbf{r,r}^{\prime };\omega \right) \equiv \left( 
\begin{array}{cc}
F_{\uparrow \uparrow }^{+}\left( \mathbf{r,r}^{\prime };\omega \right) & 
F_{\uparrow \downarrow }^{+}\left( \mathbf{r,r}^{\prime };\omega \right) \\ 
F_{\downarrow \uparrow }^{+}\left( \mathbf{r,r}^{\prime };\omega \right) & 
F_{\downarrow \downarrow }^{+}\left( \mathbf{r,r}^{\prime };\omega \right)%
\end{array}%
\right) =\int d\mathbf{r}^{\prime \prime }\widehat{G}_{11}^{\left( 0\right)
T}\left( \mathbf{r}^{\prime \prime }\mathbf{,r};-\omega \right) i\sigma
_{y}\Delta ^{\ast }\left( \mathbf{r}^{\prime \prime }\right) \widehat{G}%
_{11}^{\left( 0\right) }\left( \mathbf{r}^{\prime \prime }\mathbf{,r}%
^{\prime };\omega \right)
\end{equation*}

\bigskip

so that for the anomalous Green's functions $F_{\downarrow \uparrow
}^{+}\left( \mathbf{r,r}^{\prime };\omega \right) $ and $F_{\uparrow
\downarrow }^{+}\left( \mathbf{r,r}^{\prime };\omega \right) $ we find,
respectively:

\bigskip

\begin{equation}
F_{\downarrow \uparrow }^{+}\left( \mathbf{r,r}^{\prime };\omega \right)
=\int d\mathbf{r}^{\prime \prime }\Delta ^{\ast }\left( \mathbf{r}^{\prime
\prime }\right) \left[ G_{\uparrow \downarrow }^{\left( 0\right) }\left( 
\mathbf{r}^{\prime \prime }\mathbf{,r};-\omega \right) G_{\downarrow
\uparrow }^{\left( 0\right) }\left( \mathbf{r}^{\prime \prime }\mathbf{,r}%
^{\prime };\omega \right) -G_{\downarrow \downarrow }^{\left( 0\right)
}\left( \mathbf{r}^{\prime \prime }\mathbf{,r};-\omega \right) G_{\uparrow
\uparrow }^{\left( 0\right) }\left( \mathbf{r}^{\prime \prime }\mathbf{,r}%
^{\prime };\omega \right) \right]  \label{FDU1CC}
\end{equation}

\bigskip

\begin{equation}
F_{\uparrow \downarrow }^{+}\left( \mathbf{r,r}^{\prime };\omega \right)
=\int d\mathbf{r}^{\prime \prime }\Delta ^{\ast }\left( \mathbf{r}^{\prime
\prime }\right) \left[ G_{\uparrow \uparrow }^{\left( 0\right) }\left( 
\mathbf{r}^{\prime \prime }\mathbf{,r};-\omega \right) G_{\downarrow
\downarrow }^{\left( 0\right) }\left( \mathbf{r}^{\prime \prime }\mathbf{,r}%
^{\prime };\omega \right) -G_{\downarrow \uparrow }^{\left( 0\right) }\left( 
\mathbf{r}^{\prime \prime }\mathbf{,r};-\omega \right) G_{\uparrow
\downarrow }^{\left( 0\right) }\left( \mathbf{r}^{\prime \prime }\mathbf{,r}%
^{\prime };\omega \right) \right]  \label{FUD1CC}
\end{equation}

\bigskip

The self-consistency condition for the singlet SC order parameter, in the
imaginary (Matsubara) frequency representation, $\omega _{\nu }=\left( 2\nu
+1\right) \pi \tau \left(\tau \equiv k_{B}T/\hbar \omega _{c}\right),\nu
=0,\pm 1,\pm 2,...$:

\begin{equation}
\Delta ^{\ast }\left( \mathbf{r}\right) =-\left( \left \vert V\right \vert
/\hbar \omega _{c}\right) \left \langle \widehat{\psi }_{\downarrow
}^{\dagger }\left( \mathbf{r};\tau \right) \widehat{\psi }_{\uparrow
}^{\dagger }\left( \mathbf{r};\tau \right) \right \rangle = \left( \left
\vert V\right \vert /\hbar \omega _{c} \right) \tau \sum \limits_{\nu
=-\infty }^{\infty }F_{\downarrow \uparrow }^{+}\left( \mathbf{r,r};\omega
_{\nu }\right) ,\Delta ^{\ast }\left( \mathbf{r}\right) \equiv \Delta
_{\uparrow \downarrow }^{\ast }\left( \mathbf{r}\right) =-\Delta
_{\downarrow \uparrow }^{\ast }\left( \mathbf{r}\right)  \label{SCF}
\end{equation}%
where $V$ is the effective electron-electron interaction potential
responsible for the pairing instability. \ Note that, unlike the
dimensionless quantity $\Delta ^{\ast }\left( \mathbf{r}\right) $, both $V$
and $k_{B}T$ are expressed here in their absolute energy dimensions. \ Thus,
to leading order in $\Delta ^{\ast }\left( \mathbf{r}\right) $ Eq.\ref{SCF}
takes the form:

\begin{equation}
\Delta ^{\ast }\left( \mathbf{r}\right) =\left( \left \vert V\right \vert
/\hbar \omega _{c}\right) \int d\mathbf{r}^{\prime }\Delta ^{\ast }\left( 
\mathbf{r}^{\prime }\right) Q\left( \mathbf{r}^{\prime }\mathbf{,r}\right)
\label{SCDel}
\end{equation}%
where the kernel $Q\left( \mathbf{r}^{\prime }\mathbf{,r}\right) $ is given
by:

\begin{eqnarray}
Q\left( \mathbf{r}^{\prime }\mathbf{,r}\right) &=&\tau \sum \limits_{\nu
=-\infty }^{\infty }Q\left( \mathbf{r}^{\prime }\mathbf{,r};\omega _{\nu
}\right) ,  \notag \\
Q\left( \mathbf{r}^{\prime }\mathbf{,r};\omega _{\nu }\right)
&=&Q_{\downarrow \uparrow }^{\uparrow \downarrow }\left( \mathbf{r}^{\prime }%
\mathbf{,r};\omega _{\nu }\right) -Q_{\downarrow \downarrow }^{\uparrow
\uparrow }\left( \mathbf{r}^{\prime }\mathbf{,r};\omega _{\nu }\right)
\label{QUDDU-QUUDD}
\end{eqnarray}%
with:

\begin{eqnarray*}
Q_{\downarrow \uparrow }^{\uparrow \downarrow }\left( \mathbf{r}^{\prime }%
\mathbf{,r};\omega _{\nu }\right) &=&G_{\downarrow \uparrow }^{\left(
0\right) }\left( \mathbf{r}^{\prime }\mathbf{,r};-\omega _{\nu }\right)
G_{\uparrow \downarrow }^{\left( 0\right) }\left( \mathbf{r}^{\prime }%
\mathbf{,r};\omega _{\nu }\right) , \\
Q_{\downarrow \downarrow }^{\uparrow \uparrow }\left( \mathbf{r}^{\prime }%
\mathbf{,r};\omega _{\nu }\right) &=&G_{\downarrow \downarrow }^{\left(
0\right) }\left( \mathbf{r}^{\prime }\mathbf{,r};-\omega _{\nu }\right)
G_{\uparrow \uparrow }^{\left( 0\right) }\left( \mathbf{r}^{\prime }\mathbf{%
,r};\omega _{\nu }\right)
\end{eqnarray*}

Exploiting Eqs.\ref{gUU}-\ref{gDD} for the normal state Green's functions we
derive the following explicit expressions for the kernels:

\bigskip

\begin{eqnarray}
Q_{\downarrow \downarrow }^{\uparrow \uparrow }\left( \mathbf{r}^{\prime }%
\mathbf{,r};\omega _{\nu }\right) &=&\left( \frac{1}{2\pi }\right) ^{2}\sum
\limits_{n,n^{\prime }=0}^{\infty }\frac{\left( -i\omega _{\nu }+\mu \right)
\left( i\omega _{\nu }+\mu \right) }{\left[ \left( -i\omega _{\nu }+\mu
\right) ^{2}-n\right] \left[ \left( i\omega _{\nu }+\mu \right) ^{2}-\left(
n^{\prime }+1\right) \right] }\times  \notag \\
&&e^{i\left( x^{\prime }-x\right) \left( y+y^{\prime }\right) }e^{-\frac{1}{2%
}\rho ^{2}}L_{n^{\prime }}\left( \frac{1}{2}\rho ^{2}\right) L_{n}\left( 
\frac{1}{2}\rho ^{2}\right)  \label{QUUDD}
\end{eqnarray}

\begin{eqnarray}
Q_{\downarrow \uparrow }^{\uparrow \downarrow }\left( \mathbf{r}^{\prime }%
\mathbf{,r};\omega _{\nu }\right) &=&-\frac{1}{2}\left( \frac{1}{2\pi }%
\right) ^{2}\sum \limits_{n,n^{\prime }=1}^{\infty }\frac{1}{\left[ \left(
-i\omega _{\nu }+\mu \right) ^{2}-n\right] }\frac{1}{\left[ \left( i\omega
_{\nu }+\mu \right) ^{2}-n^{\prime }\right] }\times  \notag \\
&&e^{i\left( x^{\prime }-x\right) \left( y+y^{\prime }\right) }e^{-\frac{1}{2%
}\rho ^{2}}\rho ^{2}L_{n}^{\prime }\left( \frac{1}{2}\rho ^{2}\right)
L_{n^{\prime }}^{\prime }\left( \frac{1}{2}\rho ^{2}\right),  \label{QUDDU}
\end{eqnarray}
where $\rho = |\mathbf{r}-\mathbf{r^{\prime }}|$. Similar to the situation
in the standard, single band (with quadratic energy dispersion) 2D electron
system, the order parameter of the Landau orbital form $\Delta ^{\ast
}\left( \mathbf{r}\right) \propto e^{-2iq_{x}x}\varphi _{m}\left[ \sqrt{2}%
\left( y-q_{x}\right) \right] ,m=0,1,2,...$, is an eigenfunction of the
integral operator $\int d\mathbf{r}^{\prime }Q\left( \mathbf{r}^{\prime }%
\mathbf{,r}\right) ...,$

that is:

\begin{equation}
\int d\mathbf{r}^{\prime }Q\left( \mathbf{r}^{\prime }\mathbf{,r}\right)
\Delta ^{\ast }\left( \mathbf{r}^{\prime }\right) =A\Delta ^{\ast }\left( 
\mathbf{r}\right)  \label{EigenDel}
\end{equation}

This important result is obtained by showing that the above Landau orbital
is also an eigenfunction of the integral operators $k_{B}T\sum \limits_{\nu
=-\infty }^{\infty }\int d\mathbf{r}^{\prime }Q_{\downarrow \downarrow
}^{\uparrow \uparrow }\left( \mathbf{r}^{\prime }\mathbf{,r};\omega _{\nu
}\right)... , k_{B}T\sum \limits_{\nu =-\infty }^{\infty }\int d\mathbf{r}%
^{\prime }Q_{\downarrow \uparrow }^{\uparrow \downarrow }\left( \mathbf{r}%
^{\prime }\mathbf{,r};\omega _{\nu }\right)... $, so that, by exploiting Eqs.%
\ref{QUUDD},\ref{QUDDU}, the eigenvalue $A$ in Eq.\ref{EigenDel} can be
written in terms of the respective eigenvalues, $A_{\downarrow \uparrow
}^{\uparrow \downarrow },A_{\downarrow \downarrow }^{\uparrow \uparrow }$ ,
as:

\begin{equation}
A=A_{\downarrow \uparrow }^{\uparrow \downarrow }-A_{\downarrow \downarrow
}^{\uparrow \uparrow }  \label{A}
\end{equation}%
where:

\bigskip

\begin{eqnarray}
A_{\downarrow \uparrow }^{\uparrow \downarrow } &=&-\tau \sum \limits_{\nu
=-\infty }^{\infty }\sum \limits_{n,n^{\prime }=1}^{\infty }\frac{1}{\left[
\left( -i\omega _{\nu }+\mu \right) ^{2}-n\right] }\frac{1}{\left[ \left(
i\omega _{\nu }+\mu \right) ^{2}-n^{\prime }\right] }\times  \label{AUDDU} \\
&&\frac{1}{2}\left( \frac{1}{2\pi }\right) \int_{0}^{\infty }\rho ^{3}d\rho
e^{-\rho ^{2}}L_{n^{\prime }}^{\prime }\left( \frac{1}{2}\rho ^{2}\right)
L_{n}^{\prime }\left( \frac{1}{2}\rho ^{2}\right)  \notag
\end{eqnarray}

\bigskip

\begin{eqnarray}
A_{\downarrow \downarrow }^{\uparrow \uparrow } &=&\tau \sum \limits_{\nu
=-\infty }^{\infty }\sum \limits_{n,n^{\prime }=0}^{\infty }\frac{\left(
-i\omega _{\nu }+\mu \right) \left( i\omega _{\nu }+\mu \right) }{\left[
\left( -i\omega _{\nu }+\mu \right) ^{2}-n\right] \left[ \left( i\omega
_{\nu }+\mu \right) ^{2}-\left( n^{\prime }+1\right) \right] }\times
\label{AUUDD} \\
&&\left( \frac{1}{2\pi }\right) \int_{0}^{\infty }\rho d\rho e^{-\rho
^{2}}L_{n^{\prime }}\left( \frac{1}{2}\rho ^{2}\right) L_{n}\left( \frac{1}{2%
}\rho ^{2}\right)  \notag
\end{eqnarray}

and $\  \tau \equiv \frac{k_{B}T}{\hbar \omega _{c}}$.\bigskip

Performing the integration over $\rho $ in both Eqs. \ref{AUDDU} and \ref%
{AUUDD} our results for the eigenvalues $A_{\downarrow \uparrow }^{\uparrow
\downarrow },A_{\downarrow \downarrow }^{\uparrow \uparrow }$ take the forms:

\bigskip

\begin{equation}
A_{\downarrow \uparrow }^{\uparrow \downarrow }\equiv A_{0}=-\left( \frac{1}{%
2\pi }\right) \tau \sum \limits_{\nu =-\infty }^{\infty }\sum
\limits_{n,n^{\prime }=1}^{\infty }\frac{\left( n+n^{\prime }\right) !}{%
2^{n+n^{\prime }}n^{\prime }!n!}\frac{\left( \frac{n^{\prime }n}{n+n^{\prime
}}\right) }{\left[ \left( -i\omega _{\nu }+\mu \right) ^{2}-n\right] \left[
\left( i\omega _{\nu }+\mu \right) ^{2}-n^{\prime }\right] }  \label{AUDDUf}
\end{equation}

\bigskip

\begin{eqnarray}
A_{\downarrow \downarrow }^{\uparrow \uparrow } &=&A_{1}+\Delta A_{1},
\label{AUUDDf} \\
A_{1} &=&\left( \frac{1}{4\pi }\right) \tau \sum \limits_{\nu =-\infty
}^{\infty }\sum \limits_{n,n^{\prime }=1}^{\infty }\frac{\left( n+n^{\prime
}\right) !}{2^{n+n^{\prime }}n!n^{\prime }!}\frac{\left( -i\omega _{\nu
}+\mu \right) \left( i\omega _{\nu }+\mu \right) }{\left[ \left( -i\omega
_{\nu }+\mu \right) ^{2}-n\right] \left[ \left( i\omega _{\nu }+\mu \right)
^{2}-n^{\prime }\right] },  \label{A1} \\
\Delta A_{1} &=&\left( \frac{1}{4\pi }\right) \tau \sum \limits_{\nu
=-\infty }^{\infty }\sum \limits_{n=1}^{\infty }\frac{1}{2^{n}}\left \{ 
\frac{\left( i\omega _{\nu }+\mu \right) }{\left( -i\omega _{\nu }+\mu
\right) \left[ \left( i\omega _{\nu }+\mu \right) ^{2}-n\right] }+CC\right \}
\label{DA1}
\end{eqnarray}

\end{widetext}

Note that the pairing correlations of electrons, preserving their spin-up
projection, with electrons preserving their spin-down projection, as
expressed by $A_{\downarrow \downarrow }^{\uparrow \uparrow }$ in Eq.\ref%
{AUUDDf}, include contributions of correlations of the zero LL with nonzero
LLs, as reflected by $\Delta A_{1}$ in Eq.\ref{DA1}. On the other hand, the
pairing correlations of electrons, flipping their spin-up to spin-down
projections, due to spin orbit coupling, with electrons flipping their
spin-down to spin-up projections, as expressed by $A_{\downarrow \uparrow
}^{\uparrow \downarrow }$ in Eq.\ref{AUDDUf}, do not include any
contributions involving zero LL states.

Combining the self-consistency integral equation, Eq.\ref{SCF}, with the
eigenvalue equation (for $A=A_{\downarrow \uparrow }^{\uparrow \downarrow
}-A_{\downarrow \downarrow }^{\uparrow \uparrow }$), Eq.\ref{EigenDel}, the
former reduces to the simple algebraic equation, $1=\left \vert
V\right
\vert A $. \ Performing the summation over the Matsubara
frequencies $\omega _{\nu }$ we arrive at the following expression for $%
\left \vert V\right
\vert A$:

\begin{widetext}

\begin{eqnarray}
\left \vert V\right \vert A &=&\frac{1}{32}\frac{\lambda }{\sqrt{n_{F}}}\sum
\limits_{i,j=1}^{2}\sum \limits_{n=N_{l}^{\left( i\right) }}^{N_{u}^{\left(
i\right) }}\sum \limits_{m=N_{l}^{\left( j\right) }}^{N_{u}^{\left( j\right)
}}\frac{\left( m+n\right) !}{2^{m+n}n!m!}I_{nm}^{\left( ij\right) },
\label{Af} \\
I_{nm}^{\left( ij\right) } &=&\frac{\left[ \left( -1\right) ^{j}\sqrt{m}%
+\left( -1\right) ^{i}\sqrt{n}\right] ^{2}}{n+m}\frac{\tanh \left( \frac{\mu
+\left( -1\right) ^{j}\sqrt{m}}{2\tau }\right) +\tanh \left( \frac{\mu
+\left( -1\right) ^{i}\sqrt{n}}{2\tau }\right) }{2\mu +\left( -1\right) ^{j}%
\sqrt{m}+\left( -1\right) ^{i}\sqrt{n}},\  \ I_{00}^{\left( ij\right) }=0 
\notag
\end{eqnarray}

\end{widetext}
where:

\begin{equation*}
\lambda =\frac{\sqrt{n_{F}}\left \vert V\right \vert }{\pi a_{H}^{2}\hbar
\omega _{c}}=\left \vert V\right \vert N\left( E_{F}\right) =\left \vert
V\right \vert \left( \frac{m^{\ast }}{2\pi \hbar ^{2}}\right) ,n_{F}\equiv
\mu ^{2}
\end{equation*}%
with $N\left( E_{F}\right) =\frac{E_{F}}{2\pi \left( \hbar v\right) ^{2}}$
being the single electron density of states per spin projection per unit
area and $m^{\ast }=\frac{E_{F}}{v^{2}}$ the effective cyclotron mass at the
Fermi energy.

The different cutoff LL indices $N_{u}^{i}$ ($N_{l}^{i}$ ), indicated in Eqs.%
\ref{Af}, refer to the different branches, i.e. the conduction (positive),
or valence (negative) energy bands of the Weyl model contributing to the
pairing correlation. The different values arise due to the fact that the
cutoff is introduced to the electron energy, by the mediating
electron-phonon interaction, relative to the Fermi energy, rather than to
the branching point (zero) energy of the Weyl bands structure. \ Thus,
assuming a (Debye) cutoff energy $\hbar \omega _{D}$, we should distinguish
between two different situations. In the usual situation where $\hbar \omega
_{D}<E_{F}$, pairing takes place only in a single band, so that, e.g. for a
positive chemical potential, we find: $N_{u}^{\left( 1\right) }=\left[
n_{F}\left( 1+\gamma \right) ^{2}\right] $, $N_{l}^{\left( 1\right) }=\left[
n_{F}\left( 1-\gamma \right) ^{2}\right] $, $\ N_{u}^{\left( 2\right)
}=N_{l}^{\left( 2\right) }=0$, where $\gamma =\hbar \omega _{D}/E_{F}$. \ In
the unusual situation where the cutoff energy, $\hbar \omega _{D}>E_{F}$,
both inter and intra band pairing take place, so that the cutoff LL indices
are different for energies in the valence (V) and conduction (C) bands.
Thus, for CB pairing (corresponding to the energy denominator $2\mu -\sqrt{m}%
-\sqrt{n}$ in Eq.\ref{Af}), we have: $N_{u}^{\left( 1\right) }=\left[
n_{F}\left( 1+\gamma \right) ^{2}\right] ,N_{l}^{\left( 1\right) }=0$. For
the interband pairing (energy denominators $2\mu -\sqrt{m}+\sqrt{n}$ , or $%
2\mu +\sqrt{m}-\sqrt{n}$ in Eq.\ref{Af}) the cutoff indices are: $%
N_{u}^{\left( 1\right) }=\left[ n_{F}\left( 1+\gamma \right) ^{2}\right]
,N_{l}^{\left( 1\right) }=0$, or: $N_{u}^{\left( 2\right) }=\left[
n_{F}\left( \gamma -1\right) ^{2}\right] ,N_{l}^{\left( 2\right) }=0$,
respectively.

\begin{figure}[tbp]
\centering
\includegraphics[width=0.4\textwidth]{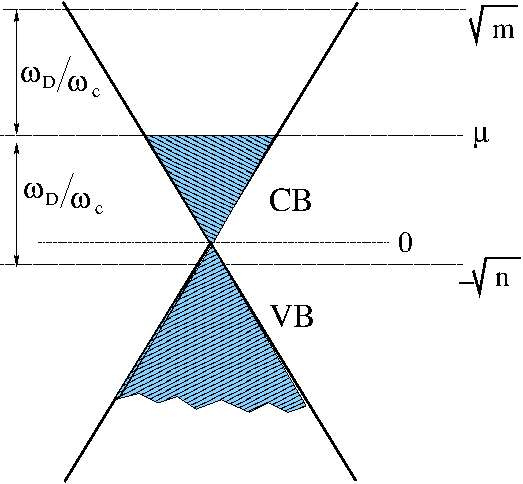}
\caption{Schematic illustration of the Weyl model bands structure for a
positive chemical potential, smaller than the cutoff energy, showing a pair
of Landau levels in both subbands at the cutoff energy measured from the
Fermi energy.}
\end{figure}

\bigskip

\section{Comparison with the standard (nonrelativistic) electron gas model:
The semiclassical approximation}

\bigskip

A useful reference model, for comparison with the 2D Weyl model developed
above, starts with a nonrelativistic electron gas, characterized by a
quadratic single-electron energy-momentum dispersion, $E=\frac{\hbar
^{2}k^{2}}{2m_{S}^{\ast }}$, with band effective mass, $m_{S}^{\ast }$ , set
equal to $\frac{1}{2}m_{0}^{\ast }=\frac{E_{F0}}{2v^{2}}$, and $E_{F0}$- the
Fermi energy in the Weyl model at a certain doping level, to be determined
in reference to a concrete experiment. Under these assumptions both the
Fermi energy, $E_{F0}$ , and the Fermi wave number, $k_{F0}$ , are the same
in both models: 
\begin{equation}
E_{F0}=\frac{\hbar ^{2}k_{F0}^{2}}{2m_{S}^{\ast }}=\hbar vk_{F0}  \label{EF}
\end{equation}%
And in a perpendicular magnetic field $\mathbf{H=}\left( 0,0,H\right) $ the
cyclotron frequency, $\omega _{c}^{S}\equiv \left( \frac{eH}{m_{S}^{\ast }c}%
\right) $, is related to the Weyl cyclotron frequency, $\omega
_{c}^{W}\equiv \frac{\sqrt{2}v}{a_{H}}$, via:

\bigskip

\begin{equation}
\omega _{c}^{W}=2\sqrt{n_{F0}}\left( \frac{eH}{m_{0}^{\ast }c}\right) =\sqrt{%
n_{F0}}\omega _{c}^{S}  \label{wc}
\end{equation}%
where in both models 
\begin{equation}
n_{F0}\equiv \frac{E_{F0}}{\hbar \omega _{c}^{S}}\equiv \left( \frac{E_{F0}}{%
\hbar \omega _{c}^{W}}\right) ^{2}=\frac{\left( a_{H}k_{F0}\right) ^{2}}{2}
\label{nF0}
\end{equation}

\bigskip

Using the set of parameters defined above, the well known expression for the
pairing energy eigenvalue obtained in the standard model takes the form:

\begin{widetext}

\begin{equation}
\left \vert V\right \vert A_{S}=\frac{1}{4}\lambda _{S}\sum
\limits_{m,n=n_{F0}\left( 1-\gamma _{0}\right) }^{n_{F0}\left( 1+\gamma
_{0}\right) }\frac{\left( m+n\right) !}{2^{m+n}n!m!}\frac{\tanh \left( \frac{%
\mu _{0}-n-1/2}{2\tau _{S}}\right) +\tanh \left( \frac{\mu _{0}-m-1/2}{2\tau
_{S}}\right) }{2\mu _{0}-n-m-1}  \label{AS0}
\end{equation}

\end{widetext}
where $\mu _{0}=n_{F0},\tau _{S}=\frac{k_{B}T}{\hbar \omega _{c}^{S}}%
,\lambda _{S}\equiv \left \vert V\right \vert \left( \frac{m_{S}^{\ast }}{%
2\pi \hbar ^{2}}\right) $, and $\gamma _{0}=\hbar \omega _{D}/E_{F0}$.

The semiclassical limit of our theory in the Weyl model is basically
established at sufficiently small magnetic fields for which the LL index at
the Fermi energy, $n_{F}$, is sufficiently large compared to unity. \ Thus,
assuming that $n_{F}\gg 1$ , we may expand the CB energy appearing in the
dominant contribution to $A$ (i.e. $I_{nm}^{\left( 11\right) }$) in Eq.\ref%
{Af} around $m=n_{F}$ , or $n=n_{F}$, e.g.: $\  \sqrt{m}\approx \mu +\frac{1}{%
2\sqrt{n_{F}}}\left( m-n_{F}\right) $, $\sqrt{n}\approx \mu +\frac{1}{2\sqrt{%
n_{F}}}\left( n-n_{F}\right) $, \ such that to leading order: $%
I_{nm}^{\left( 11\right) }\approx 4\sqrt{n_{F}}\frac{\tanh \left( \frac{%
n_{F}-m}{2\left( 2\tau \sqrt{n_{F}}\right) }\right) +\tanh \left( \frac{%
n_{F}-n}{2\left( 2\tau \sqrt{n_{F}}\right) }\right) }{2n_{F}-m-n}$, and:

\begin{widetext}
\begin{equation}
\left \vert V\right \vert A_{W}\approx \left \vert V\right \vert A_{W}^{SC}=%
\frac{1}{8}\lambda \sum \limits_{m,n=N_{l}^{\left( 1\right)
}}^{N_{u}^{\left( 1\right) }}\frac{\left( m+n\right) !}{2^{m+n}n!m!}\frac{%
\tanh \left( \frac{n_{F}-m}{2\left( 2\sqrt{n_{F}}\tau _{W}\right) }\right)
+\tanh \left( \frac{n_{F}-n}{2\left( 2\sqrt{n_{F}}\tau _{W}\right) }\right) 
}{2n_{F}-m-n}  \label{ASC}
\end{equation}
\end{widetext}
Note that, for $E_{F}=E_{F0}$, $A_{W}$ in Eq.\ref{ASC} is seen to be close
to $\frac{1}{2}A_{S}$ in Eq.\ref{AS0}, provided the dimensionless
temperature scale $\tau _{W}\equiv \frac{k_{B}T}{\hbar \omega _{c}^{W}}$ is
rescaled by the factor $2\sqrt{n_{F0}}$. \ In fact, the rescaled value, $2%
\sqrt{n_{F0}}\tau _{W}=\frac{k_{B}T}{\hbar eH/m_{0}^{\ast }c}=2\left( \frac{%
k_{B}T}{\hbar \omega _{c}^{S}}\right) =2\tau _{S}$ , is consistent with Eq.%
\ref{wc}. It should be noted that the dimensionless zero-point energy, $1/2$
in Eq.\ref{AS0}, characterizing the standard model, does not make any
difference since it can always be absorbed into the chemical potential $\mu $%
. \ The factor of $\frac{1}{2}$ between the expressions \ref{ASC} and \ref%
{AS0} is due to the spin-momentum locking, inherent to the Weyl model, and
the consequent splitting of its spectrum into positive and negative energy
subbands, as compared to the single band of the standard spectrum.

There is, however, an essential difference between the two models, and that
is the cyclotron effective mass in the Weyl model is a function of the Fermi
energy, whereas in the standard model it is a constant. The above comparison
is, therefore, drastically modified in the ultimate quantum limit, when
together with the doping level, the Fermi energy tends to zero, and the
prefactor $\frac{\lambda }{\sqrt{n_{F}}}$ in Eq.\ref{Af} nominally diverges
as $n_{F}\rightarrow 0$. \ The vanishing of $\lambda $ in the Weyl model
with $E_{F}$ through the cyclotron effective mass, evidently removes this
divergency, yielding: $\frac{\lambda }{\sqrt{n_{F}}}\rightarrow \frac{1}{%
\sqrt{2}\pi }\frac{\left \vert V\right \vert /\hbar v}{a_{H}}%
,E_{F}\rightarrow 0$ .

It will be, therefore, helpful to extend the reference model expressed in Eq.%
\ref{AS0} for varying values of $E_{F}$, to account for the dependence of
the parameters $n_{F}$ and $m^{\ast }$ in the Weyl model on $E_{F}$. \ This
can be done by replacing $n_{F0}=\mu _{0}$ in Eq.\ref{AS0} with $n_{F}$,
defined in the Weyl model by: \bigskip 
\begin{equation}
n_{F}\equiv \left( \frac{E_{F}}{\hbar \omega _{c}^{W}}\right) ^{2}=\frac{1}{2%
}\left( a_{H}k_{F}\right) ^{2}  \label{nF}
\end{equation}%
so that:

\begin{widetext}

\begin{equation}
\left \vert V\right \vert A_{S}\rightarrow \frac{1}{4}\lambda _{S}\sum
\limits_{m,n=n_{F}\left( 1-\gamma \right) }^{n_{F}\left( 1+\gamma \right) }%
\frac{\left( m+n\right) !}{2^{m+n}n!m!}\frac{\tanh \left( \frac{n_{F}-n-1/2}{%
2\tau _{S}}\right) +\tanh \left( \frac{n_{F}-m-1/2}{2\tau _{S}}\right) }{%
2n_{F}-n-m-1}  \label{AS}
\end{equation}%
\end{widetext}
where \ $\tau _{S}=\sqrt{n_{F0}}\tau _{W}$\ and $\gamma =\hbar \omega
_{D}/E_{F}$. \  \ The standard coupling constant, $\lambda _{S}$ , is defined
by fixing the value of the cyclotron mass at $m_{S}^{\ast }$ (i.e. at a
certain value of the Fermi energy $E_{F0}$): $\lambda _{S}\equiv \left \vert
V\right \vert \left( m_{S}^{\ast }/2\pi \hbar ^{2}\right) =\frac{1}{2}%
\lambda _{0}$, so that the prefactor in Eq.\ref{Af} is rewritten in a form
showing its independence of $E_{F}$:

\begin{equation}
\frac{\lambda }{\sqrt{n_{F}}}=\frac{\lambda _{0}}{\sqrt{n_{F0}}}=2\frac{%
\lambda _{S}}{\sqrt{n_{F0}}}  \label{pref}
\end{equation}

Using this expression in Eq.\ref{Af}, together with the semiclassical
approximation that yields Expression \ref{ASC}, the pre-factor, $\lambda /8$%
, in the latter becomes: $\left( \lambda _{S}/4\right) \left( \frac{k_{F}}{%
k_{F0}}\right) $, in full agreement with Eq.\ref{AS} at the reference point $%
k_{F}=k_{F0}$. For doping levels away from the reference point, i.e. for $%
k_{F}\neq k_{F0}$, one finds the simple relation: 
\begin{equation}
A_{W}^{SC}=\left( \frac{k_{F}}{k_{F0}}\right) A_{S}  \label{ASCvsAS}
\end{equation}

\begin{widetext}
\onecolumngrid
\begin{figure*}[t]
\centering
{\label{fig:a}}{\includegraphics[width=0.45\textwidth]{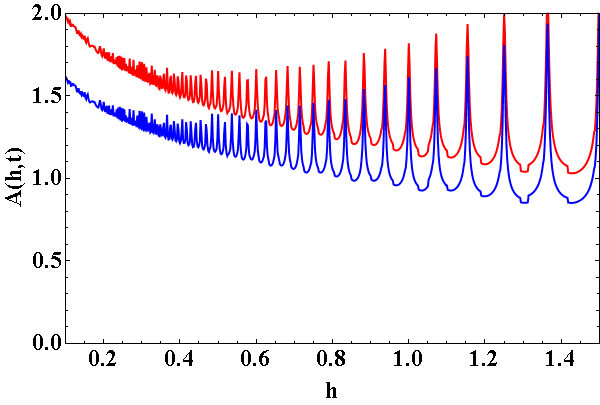}} {\label%
{fig:b}}{\includegraphics[width=0.45\textwidth]{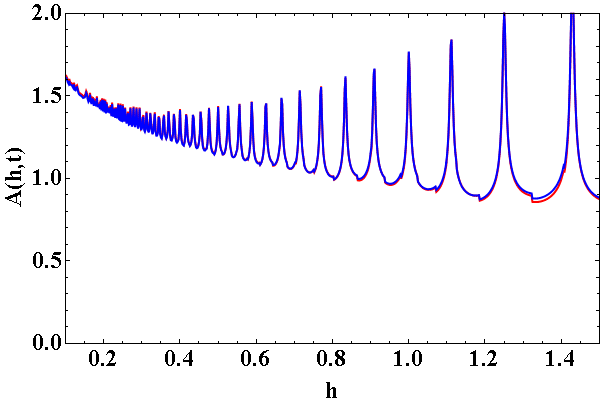}} {\label{fig:c}}{%
\includegraphics[width=0.45\textwidth]{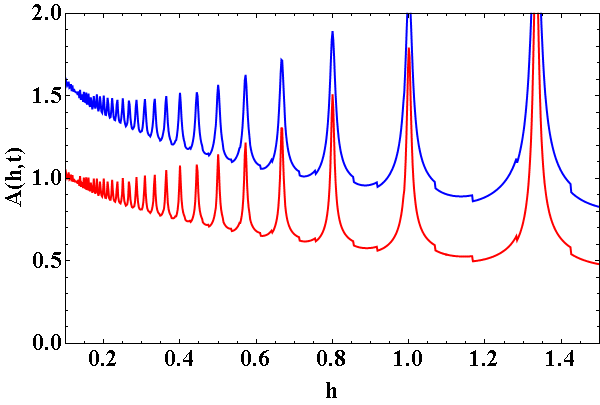}} {\label{fig:d}}{%
\includegraphics[width=0.45\textwidth]{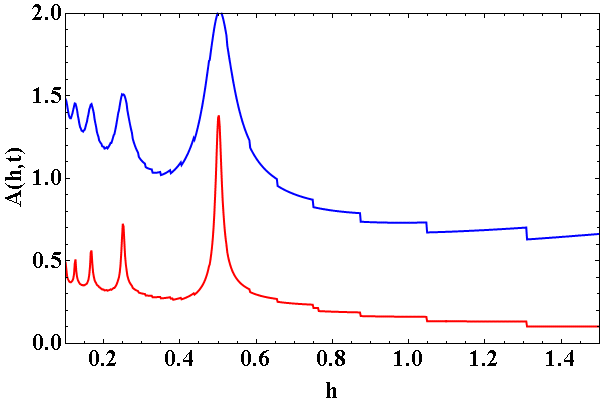}}
\caption{Pairing condensation energy eigenvalue, $A$, as a function of
field, $h\equiv H/H_{0}$ , at temperature $t\equiv T/T_{0}=0.01$, calculated
for the Weyl model, Eq.\protect \ref{Af} (red curves), and for the extended
standard model, Eq.\protect \ref{AS} (blue curves), at various values of $%
\widetilde{n}_{F\text{ }}$($=15$ (a),$10$ (b), $5$ (c), $0.5$ (d)). The
reference parameters, $H_{0},T_{0}$ , were selected in accord with the
experiment, as discussed in the text. The cutoff was selected at: $\hbar 
\protect \omega _{D}/E_{F0}=0.5$. Note that for $\widetilde{n}_{F\text{ }}<5$
the cutoff energy $\hbar \protect \omega _{D}>E_{F}$.}
\label{gene.fig1}
\end{figure*}
\twocolumngrid
\end{widetext}

\section{Mapping between the two models and their comparison with experiment}

Experimental evidence for the existence of strong type-II superconductivity
in a surface state of a topological insulator under a strong magnetic field {%
can} be found in results of transport, magnetic susceptibility, de Haas van
Alphen ({dHvA}) oscillations and scanning tunnelling spectroscopy
measurements, reported recently on Sb$_{2}$Te$_{3}$ \cite{Zhao15}. Using a
simple s-wave BCS model, similar to the standard model described in Sec.3,
with the experimentally observed {dHvA frequency,} $F_{0}=36.5$ T (implying $%
n_{F0}\left( H\right) =\frac{F_{0}}{H}$), {and cyclotron mass} $m_{0}^{\ast
}=0.065m_{e}$, it was shown in \cite{ZDMPRB17} that such an unusual SC state
can exist only in the {strong coupling superconductor limit. In particular,
the zero field limit of the self-consistent order parameter amplitude,} $%
\Delta _{SC}\left( n_{F0}\rightarrow \infty \right) \rightarrow \hbar \omega
_{D}/\sinh \left( 1/\lambda _{0}\right) $, calculated in \cite{ZDMPRB17},
was found, for $\lambda _{0}=1$ and $\hbar \omega _{D}=0.25E_{F0}$, to
basically agree with {the spatially average SC energy gap, derived from the
STS measurements} (i.e. $\simeq 13$ meV) \cite{Zhao15}, whereas the LL
filling factor, calculated at the semiclassical $H_{c2}$ ( $n_{F0}\approx 14$%
), was found to agree with the experimentally determined field of the
resistivity onset downshift $H_{R}$ ($\sim 2.5$ T, $n_{F0}\sim 14$) \cite%
{Zhao15}. \ 

Such an agreement, between the standard model, outlined in Sec.3, and the
experiment reported in \cite{Zhao15}, seems to imply that the peculiar
features of the helical surface state bands structure distinguishing the 2D
Weyl Fermion gas model from the standard model, are irrelevant in
constructing its high-fields SC state, except for a single parameter:- its
unusual cyclotron effective mass, which can be dramatically modified upon
variation of the chemical potential (e.g. by doping or by changing the gate
voltage). The analysis presented in Sec.3 supports this conclusion for
carrier densities and magnetic fields in the semiclassical limit. \ 

Here we study the relationships between the Weyl model and the extended
standard model, described above, in the general parameters range, finding
conditions for a complete mapping between the two models, and searching for
physical situations in which they are qualitatively distinguishable. In
Fig.2 we plot results of the pairing eigenvalue $A$, calculated within both
the Weyl and the extended standard models, as a function of the reduced
magnetic field, $h\equiv H/H_{0}$, for various values of Fermi energy $E_{F}$%
. \ The temperature was selected sufficiently small to unfold the quantum
oscillations associated with the Landau quantization.

\begin{figure}[t]
\centering
\includegraphics[width=0.4\textwidth]{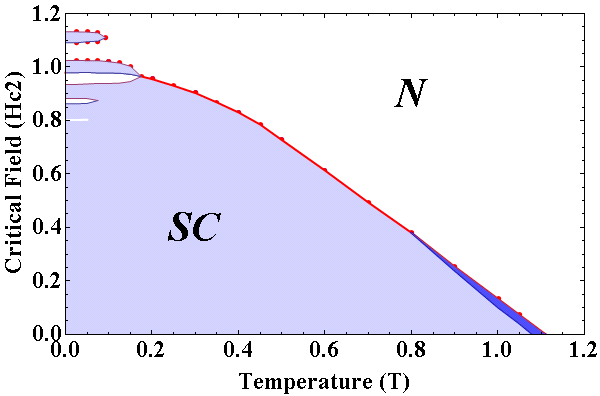}
\caption{H-T Phase diagram, obtained by solving the self-consistency
equation for both models at the reference point: $\widetilde{n}_{F}=%
\widetilde{n}_{F0}=10$, on the basis of the reference parameters $%
H_{0},T_{0} $ , as discussed in the text.The cutoff was selected at: $\hbar 
\protect \omega _{D}/E_{F0}=0.5$. Deviations are seen only around the dark
blue area, where the Weyl phase boundary is slightly above the standard one.
\ Mutual reentrances of the SC and N phases, due to strong
magneto-oscillations effect, are seen around the upper-left corner of the
phase diagram. }
\label{Fig3}
\end{figure}

Selecting for the reference parameters the values extracted from the
transport and magneto-oscillations measurements \cite{Zhao15}, as described
above: $F_{0}=36.5T,H_{0}=2.5T$, $m_{0}^{\ast }=0.065m_{e}$, and from the
magnetic susceptibility measurements \cite{Zhao15} the value: $T_{0}=100K$,
we define the dimensionless reference parameters: $\widetilde{\tau }%
_{S}\equiv \left( k_{B}T_{0}/\hbar \widetilde{\omega }_{c}^{S}\right) $ and $%
\widetilde{\tau }_{W}\equiv \left( k_{B}T_{0}/\hbar \widetilde{\omega }%
_{c}^{W}\right) $, where $\widetilde{\omega }_{c}^{S}\equiv \left(
eH_{0}/m_{S}^{\ast }c\right) $ and $\widetilde{\omega }_{c}^{W}\equiv \left( 
\sqrt{2}v/a_{H_{0}}\right) $, so that: $\tau _{S}=\widetilde{\tau }%
_{S}\left( t/h\right) $ and $\tau _{W}=\widetilde{\tau }_{W}\left( t/\sqrt{h}%
\right) $. \ The two scales are therefore related via: $\widetilde{\tau }%
_{S}=\left( \widetilde{n}_{F0}\right) ^{1/2}\widetilde{\tau }_{W}$ , where $%
\widetilde{n}_{F0}\equiv \left( a_{H_{0}}k_{F0}\right) ^{2}/2\approx 10$.\  \ 

The eigenvalues $A_{W},A_{S}$, plotted in Fig.2 as functions of $h$, for
various values of $\widetilde{n}_{F}\equiv \left( a_{H_{0}}k_{F}\right)
^{2}/2$, show at $\widetilde{n}_{F}=\widetilde{n}_{F0}$ complete agreement
between the two models, including the fine structure of the quantum
oscillations, provided $\widetilde{\tau }_{S}$ is re-scaled to $2\times
\left( \widetilde{n}_{F0}\right) ^{1/2}\widetilde{\tau }_{W}$, as found in
the semiclassical approximation, Eq.\ref{ASC}. Under these conditions,
solutions of the self consistency equation, $1=\left \vert V\right \vert A$,
for both models, yield nearly identical results for the H-T phase diagrams,
as shown in Fig.3, except for a small deviation in the low fields region,
due to the different ultraviolet divergency predicted by the two models. The
two intersection points of the phase boundary with the axes, shown in Fig.3,
are seen to be close to $t=1$ and $h=1$, thus indicating that the calculated 
$H_{c2}\left( T\rightarrow 0\right) $ and $T_{c}\left( H\rightarrow 0\right) 
$ values are close to the values of $H_{0}$ and $T_{0}$, respectively.

For values of $\widetilde{n}_{F}$ away from $\widetilde{n}_{F0}$ the
baseline of $A_{W}$ is shifted with respect to that of $A_{S}$, depending on
wether $\widetilde{n}_{F}>\widetilde{n}_{F0}$ (shift up), or $\widetilde{n}%
_{F}<\widetilde{n}_{F0}$ (shift down), thus reflecting the dependence of the
pairing correlation in the Weyl model on the carrier density. This behavior
is consistent with the relation \ref{ASCvsAS} derived in the semiclassical
limit. The oscillatory patterns remain nearly the same, except for slight
relative narrowing of the Weyl peaks upon decreasing $\widetilde{n}_{F}$,
which becomes quite significant in the quantum limit, e.g. at $\widetilde{n}%
_{F}=0.5$ in Fig.2d. It is also remarkable that in the ultimate quantum
limit, i.e. when $\widetilde{n}_{F}\rightarrow 0$, the pairing correlation
in the Weyl model, despite its vanishing normal electron density of states
at the Fermi energy, does not vanish.

\section{Conclusion}

In this paper we have developed a Nambu-Gorkov Green's function approach to
strongly type-II superconductivity in a 2D spin-momentum locked (Weyl) Fermi
gas model at high perpendicular magnetic fields in order to study the
transition to high field surface superconductivity observed recently on the
topological insulator Sb$_{2}$Te$_{3}$\cite{Zhao15}. We have found that, for
LL filling factors larger than unity, superconductivity in such a 2D Weyl
Fermion gas can be mapped onto the standard 2D electron (or hole) gas model,
having the same Fermi surface parameters, but with a cyclotron effective
mass, $m^{\ast }=E_{F}/2v^{2}$ , which could be dramatically reduced below
the free electron mass, $m_{e}$, by manipulating the doping level, or the
gate voltage. \ Our calculations for Sb$_{2}$Te$_{3}$ show that the SC
helical surface state reported in \cite{Zhao15} was in the moderate
semiclassical range ($n_{F}\geq 10$), so justifying the mapping with the
standard model. They reveal a very unusual, strong type-II superconductivity
at low carrier density and small cyclotron effective mass, $m^{\ast }={0.065}%
m_{e}$, which can be realized only in the strong coupling ($\lambda \sim 1$)
superconductor limit\cite{ZDMPRB17}. Further reduction of the carrier
density in such a system could yield an effective cyclotron energy
comparable to or larger than the Fermi energy, LL filling factors smaller
than unity, and cutoff energy larger than the chemical potential, resulting
in significant deviations from the predictions of the standard model.

Note, however, that for such a dilute fermion gas system the simple mean
field BCS theoretical framework of superconductivity, exploited in this
paper, should be drastically revised, particularly due to the neglect of
both phase and amplitude fluctuations of the SC order parameter \cite%
{Emery-Kivel-Nat95}, and to the breakdown of the adiabatic approximation in
the electron phonon system \cite{GorkovPRB16}. Several recent reports on
superconductivity in very dilute fermion gas systems, such as that found in
compensated semimetallic FeSe \cite{KasaharaNCom16}, or in the large-gap
semiconductor SrTiO$_{3}$ \cite{EdgePRL15}, have drawn much attention to
fluctuation superconductivity beyond the Gaussian approximation, which could
lead to crossover between weak-coupling BCS and strong-coupling
Bose-Einstein condensate limits \cite{RanderiaBCS-BEC14}. In the presence of
strong magnetic fields the situation is further complicated due to complex
interplay between vortex and SC amplitude fluctuations \cite{ManivPRB06}.

\end{document}